\documentclass{PoS}
\usepackage{graphicx}
\usepackage{amsmath,amsfonts}

\newcommand{\ev}[1]{\langle #1 \rangle}

\title{Interface tension of the 3d 4-state Potts model using the Wang-Landau algorithm}

\ShortTitle{Interface tension using the Wang-Landau algorithm}

\author{\speaker{Ari Hietanen}\\
        CP3-Origins and the Danish Institute for Advanced Study DIAS, University of Southern Denmark, Campusvej 55, DK-5230 Odense M, Denmark\\
        E-mail:~\email{hietanen@cp3-origins.net}}

\author{Biagio Lucini\\
  College of Science, Swansea University, Singleton Park, Swansea SA2 8PP, UK
  E-mail:~\email{b.lucini@swansea.ac.uk}}

\abstract{We study the interface tension of the 4-state Potts model in
  three dimensions using the Wang-Landau algorithm. The interface
  tension is given by the ratio of the partition function with a twisted boundary condition in one
  direction and periodic boundary conditions in all other directions
  over the partition function with periodic boundary conditions in all
  directions. With the Wang-Landau algorithm we can explicitly
  calculate both partition functions and obtain the result for all
  temperatures. We find solid numerical evidence for perfect
  wetting. Our algorithm is tested by calculating thermodynamic
  quantities at the phase transition point.\\\\ 
Preprint: CP3-Origins-2011-038 and DIAS-2011-31 }

\FullConference{ The XXIX International Symposium on Lattice Field Theory - Lattice 2011\\
July 10-16, 2011\\
Squaw Valley, Lake Tahoe, California}

\begin{document}

\section{Introduction}

The  interface free energy on a $D$-dimensional lattice, with size $L_1 \times L_2\times \dots \times L_D \equiv L_1 \times V_{D-1}$, is obtained from the ratio of two partition functions
\begin{equation}
  F_s = -\log \frac{Z_a}{Z_p}-\log L_1,
\end{equation}
where $Z_p$ is a partition function with periodic boundary conditions
and $Z_a$ has periodic boundary conditions in $D-1$ directions and a
twisted boundary condition, which develop a topological excitation, in
the direction labelled by 1. In other words, the topological
excitation forced by the twisted boundary condition induces an interface in to the system, and the free energy of the interface is obtained by subtracting the free energy of the bulk ($-\log Z_p$) out of it. The $\log L_1$ term has been subtracted since the interface can form at any of the $L_1$ points. The interface tension is then 
\begin{equation}
  \sigma = \lim_{L_i\rightarrow\infty} \frac{F_s}{V_{D-1}}.
\end{equation}

The difficulty in calculating the interface tension arises from the
fact that the partition functions are hard to calculate using
traditional simulation methods, since they have exponential
fluctuations in volume
\cite{deForcrand:2000fi,deForcrand:2005rg}. Therefore, the standard
approach has been to relate the interface tension to quantities that are easier to determine with Monte Carlo simulations \cite{Kajantie:1989iy}. However, these methods generate large systematic and/or statistical errors.

In \cite{Hietanen:2011jy} we have proposed a new method to calculate
interface tensions. It is based on the Wang-Landau \cite{Wang:2001ab}
algorithm. This algorithm enables us to calculate numerically the
density of states of the system, and from this the partition function
can be obtained as a numerical integral. As a test system we used the 3d
4-state Potts model. We were also able to provide solid numerical
evidence for perfect wetting. 

\section{Wang-Landau algorithm}
Consider a system with partition function
\begin{equation}
  Z = \sum_{\{i\}} \exp(-\beta E_i) \equiv \sum_i g(E_i) \exp(-\beta E_i).
\end{equation}
where $\{i\}$ is a generic configuration and $g(E_i)$ the density of
states with energy $E_i$. The Wang-Landau algorithm~\cite{Wang:2001ab}
estimates directly the density of states $g(E_i)$. 
The algorithm is based on a simple observation: If the state of a
system is changed randomly, and the probability to visit a given
energy level $E$ is $1/g(E)$, a random walk in $E$ generates a flat
distribution across the energy levels.

The algorithm consists of following steps
\begin{itemize}
\item[{\bf 1.}] Start with any lattice configuration and a nonzero  density of states $g(E)$.
\item[{\bf 2.}] Change a random site to a random value.
\item[{\bf 3.}] Accept the change with a probability 
  \begin{equation}
    P = \min \left\{\frac{g(E_{\rm old})}{g(E_{\rm new})},1\right\}
  \end{equation}
  and if the change is rejected set $E_{\rm new} = E_{\rm old}$.
\item[{\bf 4.}] Update $g(E_{\rm new})$ and a histogram $H(E_{\rm new})$
  \begin{align}
    g(E_{\rm new})&\rightarrow f g(E_{\rm new})  \ , \nonumber\\
    H(E_{\rm new})&\rightarrow H(E_{\rm new})+1 \ .
  \end{align}
\item[{\bf 5.}] Set $E_{\rm old} = E_{\rm new}$ and repeat from 2 until $g(E)$ has converged. Then reduce $f\rightarrow\sqrt{f}$ and set $H(E)$ to zero.
\item[{\bf 6.}] Repeat from 2 until $f$ reaches some small enough value.
\end{itemize}

The algorithm satisfies detailed balance only in the limit $f\rightarrow1$. In more detail, the transition probabilities satisfy
\begin{equation}
  \frac{1}{g(E)}p(E\rightarrow E') =\left\{ 
  \begin{array}{lcl}
    \frac{1}{fg(E')}p(E'\rightarrow E), & {\rm if} &  g(E)\ge fg(E')\\
    \frac{1}{g(E)}p(E'\rightarrow E), & {\rm if } & g(E')<g(E)< fg(E')\\
    \frac{1}{g(E')}p(E'\rightarrow E), &  {\rm if} & g(E)<g(E')
  \end{array}
  \right..
\end{equation}
At $f=1$ we obtain the detailed balance condition
\begin{equation}
 \frac{1}{g(E)}p(E\rightarrow E') = \frac{1}{g(E')}p(E'\rightarrow E).
\end{equation}
Hence, with $f\ne1$ the algorithm produces a $g(E)$ that favors the large values over the small ones. However, in practice this does not seem to generate problems as long as the number of iterations is large enough, s.t., the statistical errors are larger than the systematic ones.

As a convergence criterion we follow \cite{Zhou:2005aa}, and require
that each energy level is visited  $1/\sqrt{\ln f}$ times since after
this statistical errors are not expected to decrease. This criterion
might suffer from systematic errors, which can lead the system to convergence
to a wrong density of states \cite{Morozov:2007aa,Morozov:2009aa}. A
safer criterion would be to require  $1/\ln f$ visits to each
energy level. However, for the level of accuracy requested in our
simulations, the results obtained with these two criteria are
identical within the statistical errors. See next section.

\section{Algorithm testing}
As a test case we use 4-states Potts model in 3d. The Hamiltonian of the model is
\begin{equation}
  H = 2 J \sum_{\langle ij\rangle}\left(\frac14-\delta_{q_i,q_j}\right),
\end{equation}
where $J$ is the strength of the interaction, $q_i \in [0,3]$, $\delta_{q_i,q_j}$ is the Kronecker
delta function of the spin variables $q_i,q_j$ on neighbor sites $i,j$ and the
sum $\langle ij \rangle$ is over nearest neighbors.

To obtain statistical error estimates for measured quantities we
perform 20 separate runs with twisted and untwisted boundary
conditions. For each run we do 23 iterations, meaning that $f_{\rm
  min}\approx 1 + 1.3\times10^{-7}$ during the last step (we start
by setting $f = 3$). The simulations are done on lattice sizes $L=16,20,24,28,32,40,48$, and $56$.

\begin{figure}
  \begin{center}
    \includegraphics[width=0.9\textwidth]{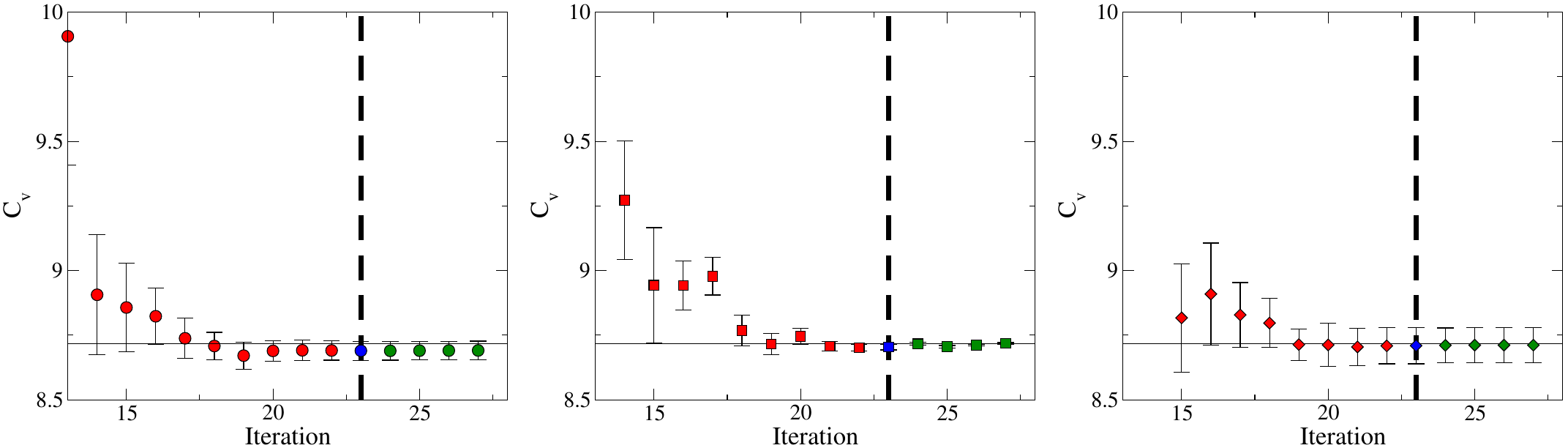}
    \caption{Specific heat calculated using three different
      convergence requirements: flat histogram (left), $1/\ln(f)$
      visits required (middle) and $1/\sqrt{\ln f}$ visits required
      (right)\label{fig_sheatit}. The solid line is the result of the middle panel after 27 iterations.}
  \end{center}
\end{figure}

\begin{figure}
  \begin{center}
    \includegraphics[width=0.45\textwidth]{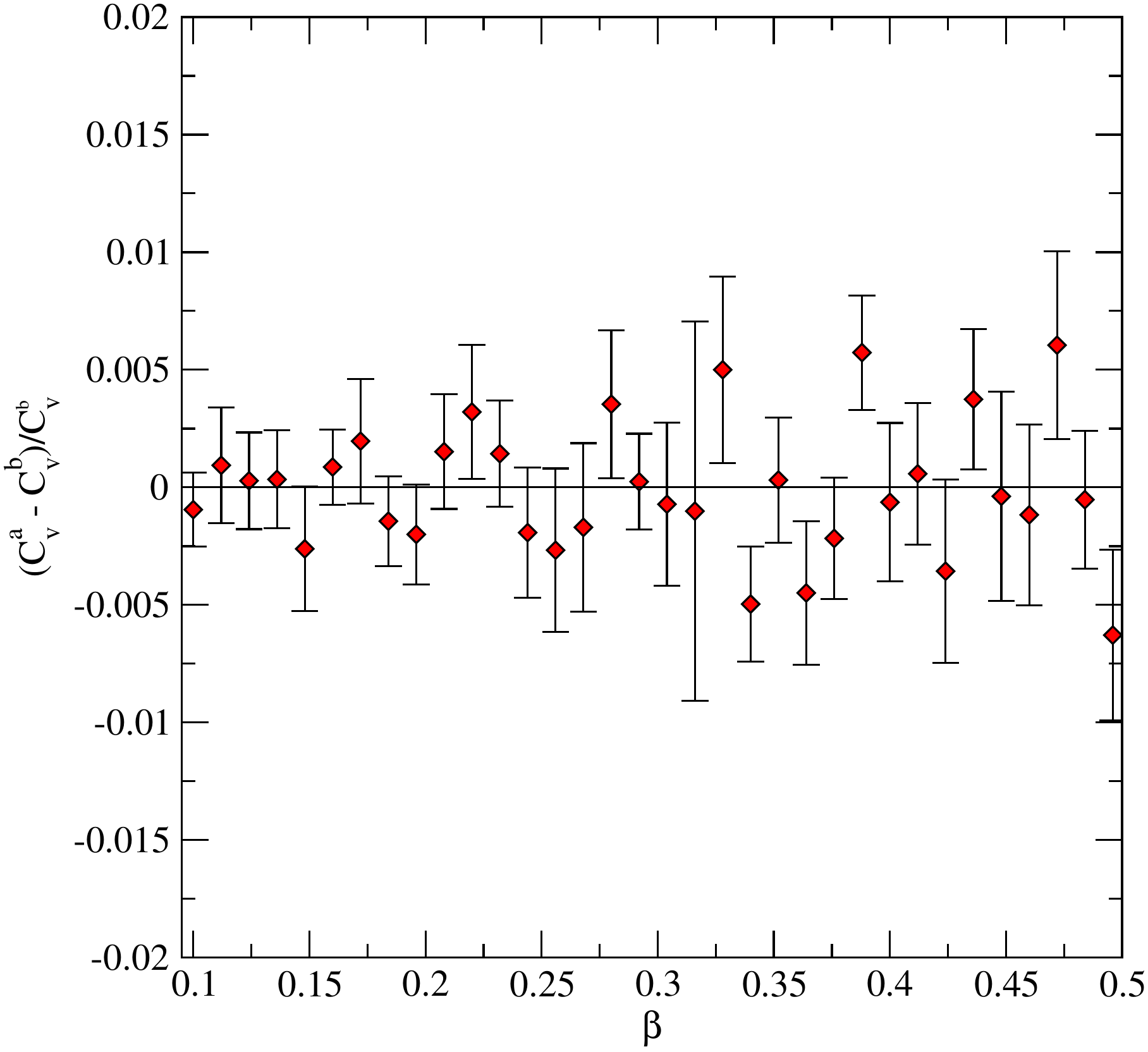}
    \hspace{20pt}
    \includegraphics[width=0.45\textwidth]{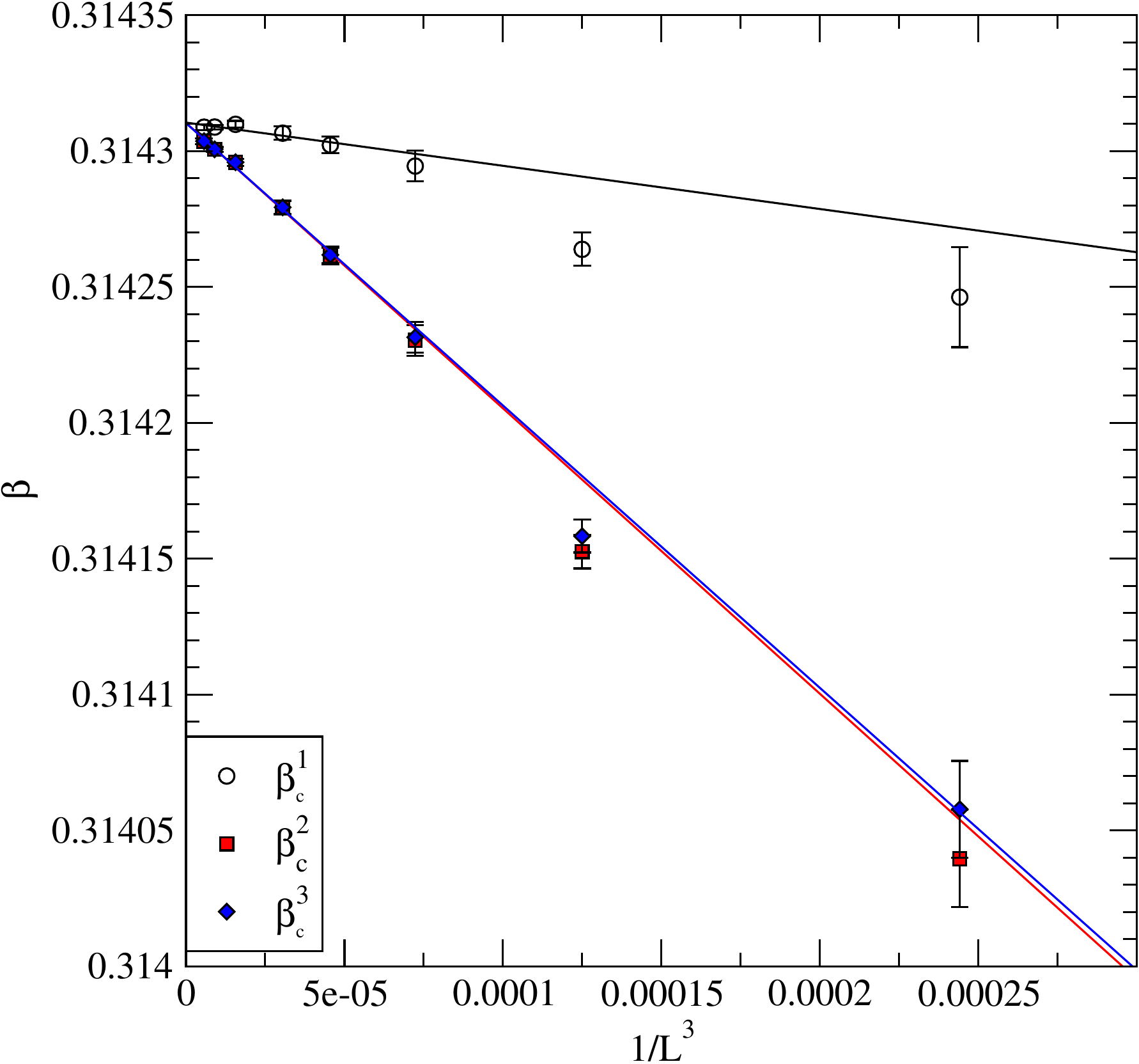}
    \caption{Left: Normalized difference between the specific heat
      obtained with two different convergence criteria. The label $a$
      indicates the criterion requiring $1/\sqrt{\ln f}$ visits at
      iteration 23 and $b$ $1/\log f$ visits at iteration 27 . Right:
      Transition temperature calculated with three different
      methods.\label{fig_sheatdiff}} 
  \end{center}
\end{figure}

To test the convergence criteria and  the systematic errors arising from the update, we measure the specific heat
\begin{equation}
  C = \frac{\beta^2}{L^3}\left(\ev{E^2}-\ev{E}^2\right) \label{sHeat}.
\end{equation}
We have used three different convergence criteria. The first requires
flat histogram, the second $1/\ln(f)$ visits and third $1/\sqrt{\ln(f)}$
visits to each energy level. In Fig.~\ref{fig_sheatit} we have plotted
the specific heat calculated with the different convergence criteria
as a function of the number of iterations. At about iteration 20 all
the three criteria provide convergence to the same value. In the left
panel of Fig.~\ref{fig_sheatdiff}, we have plotted the normalized
difference of the specific heat obtained with the third criterion at
iteration 23 and the specific heat obtained with the second criterion at
iteration 27. The differences are completely dominated by statistical
errors, and hence we do not expect to have systematic errors due to
the finite $f$ or wrong convergence of $g(E)$. 

\begin{table}
  \begin{center}
  \begin{tabular}{|c|c|}
    \hline
    $\beta^1_c$ & 0.3143104(9) \\
    $\beta^2_c$ & 0.3143103(9) \\ 
    $\beta^3_c$ & 0.3143102(9) \\ 
    \hline
    $\beta_c$ & 0.3143103(9)  \\
    $\beta_{\rm MM}$ & 0.3143103(5)  \\
    $\beta_{\rm BBD}$ & 0.3143041(17)  \\
    \hline
  \end{tabular}
  \hspace{60pt}
  \begin{tabular}{|c|c|}
    \hline
    $\Delta e$ & 1.16454(16)\\
    $\Delta e_{\rm MM}$ & 1.16492(12) \\
    $\Delta e_{\rm BBD}$ &1.16294(61) \\
    \hline
  \end{tabular}
  \caption{Comparison of our results with previous studies for the
    transition temperature (left) and the latent heat (right). The result with label MM are from 
\cite{MartinMayor:2006gx} and BBD from  \cite{Bazavov:2008qg}.
\label{table_tcext}}
  \end{center}
\end{table}

To additionally test our algorithm we compute the critical temperature
and latent heat and compare them with the results of
\cite{Bazavov:2008qg} and \cite{MartinMayor:2006gx}. Following
\cite{Bazavov:2008qg}, we use three different definitions for the
critical $\beta$ at finite volume. The first, $\beta_c^1$, is defined
to be the value for which the canonical distribution $P(E,\beta) =
g(E)e^{-\beta E}$ has two equal maxima, the second, $\beta_c^2$, is
the position of the central energy of the latent heat, and the third,
$\beta_c^3$, is the position of the maximum of the specific heat
Eq.~(\ref{sHeat}). After extrapolating to infinite volume these
definitions agree with each other. See Fig.~\ref{fig_sheatdiff}, right. 

The latent heat $\Delta e$ can be obtained from the maxima of specific heat
\begin{equation}
  C_{\rm max}(L) = c + \frac14 (\beta_c)^2 (\Delta e)^2 L^3.
\end{equation}

Our results and those of \cite{Bazavov:2008qg,MartinMayor:2006gx} are
collected in Table~\ref{table_tcext}. The discrepancy with~\cite{MartinMayor:2006gx} is less than two
standard deviations for all the quantities. The agreement is not as good with
\cite{Bazavov:2008qg}, but in this work smaller lattices are used.

\section{Interface tension}

\begin{figure}
  \begin{center}
    \includegraphics[width=0.4\textwidth]{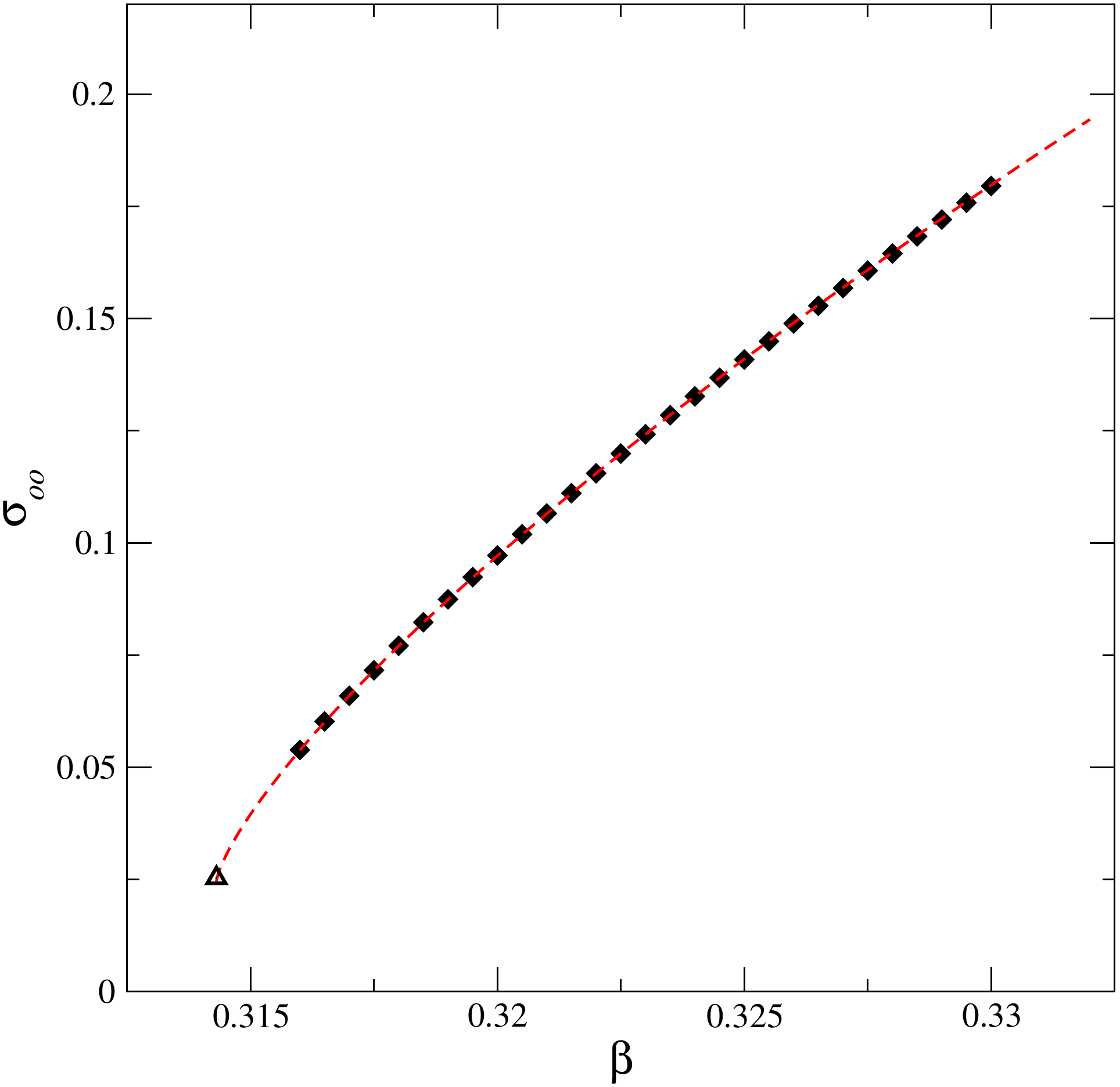}
    \includegraphics[width=0.45\textwidth]{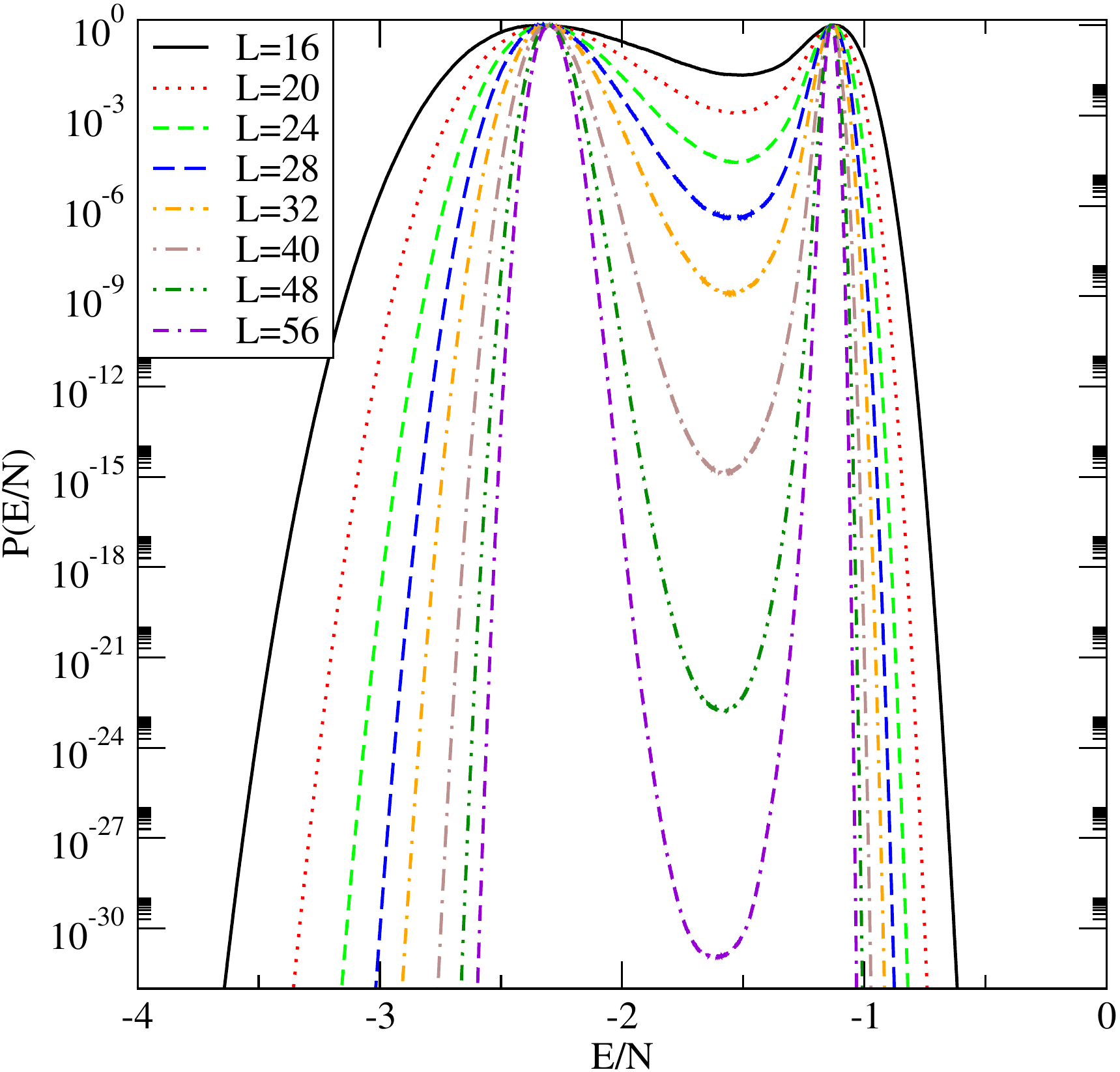}
    \caption{Left: The order-order interface tension as a function of $\beta$. Right: The probability distribution at the critical temperature. The maxima are normalized to one. \label{fig_sigmavsbeta}}
  \end{center}
\end{figure}

We consider two different kinds of interface. The first,
\emph{order-order interface}, is between two regions of space that are
in two different ordered  vacua. Near the critical temperature, the interface tension assumes the form (see \cite{Billo:2006zg,Caselle:2007yc})
\begin{equation}
  \sigma_{\rm oo}(L) = \sigma_{\rm oo} + \frac{c_2}{L^2} + \frac{c_4}{L^4} \dots \nonumber,
  \label{sigmaooL}
\end{equation}
from which the asymptotic behavior can be extracted. In our simulations we have truncated the series to $\mathcal{O}(L^{-4})$, since the corrections from higher order terms are same the size of the statistical errors. For the fit we use only the points for which
\begin{align}
  L \sqrt{\sigma_{\rm oo}} \ge 6  \nonumber \\
  \frac{1}{\sqrt{\sigma_{\rm oo}}} \ge 3.
\end{align}
The former condition is to reduce finite size effects and the latter
one is dictated by the fact that Eq.~(\ref{sigmaooL}) is only expected to hold at large distances.  

Near $\beta_c$, the behavior of $\sigma_{\rm oo}$ can be parametrized as
\begin{equation}
  \sigma_{\rm oo}(\beta) = \sigma_{\rm oo}(\beta_c) + a(\beta-\beta_c)^\rho.
\end{equation}
This functional form represents the data well. A fit with 9 degrees of freedom has $\chi^2/9=0.11$. The values are $\sigma_{oo} = 0.0249(6)$ and $\rho=0.76(4)$. See the right panel of Fig.~\ref{fig_sigmavsbeta}.

The other interface, the \emph{order-disorder interface}, is an
interface that separates an ordered state from the disordered
state. It is defined only at the critical temperature for systems with a first order phase transition. The order-disorder interface can be calculated from the probability distribution. Namely, if $P_{\rm max}$ is the peak of the histogram when the two maxima have equal height, and $P_{\rm min}$ is the minimum between the peaks, the tension is given by (see \cite{Lee:1990ti,Berg:1992qua})
\begin{equation}
  \label{ooestimator}
  2\sigma_{\rm od} = \frac{1}{L^2}\log\left(\frac{P_{\rm max}}{P_{\rm
        min}}\right) \ .
\end{equation}
Our data for $P$ are displayed in the right pane of Fig.~\ref{fig_sigmavsbeta}.

To obtain the asymptotic form we use a fitting function of the form.
\begin{equation}
  2\sigma_{\rm od} = \frac{\log L}{L^2} - 2 \sigma_{\rm od} + \frac{c_2}{L^2} + \frac{c_4}{L^4}.
  \label{fit_sigmado}
\end{equation}
A fit of the date according to Eq.~(\ref{fit_sigmado}) yields
$2\sigma_{do} = 0.0252(4)$. The insertion of a term $c_3/L^3$ in the
fitting function, justified by the use of the estimator~(\ref{ooestimator}), gives a
compatible result.

\begin{figure}
  \begin{center}
    \includegraphics[width=0.4\textwidth]{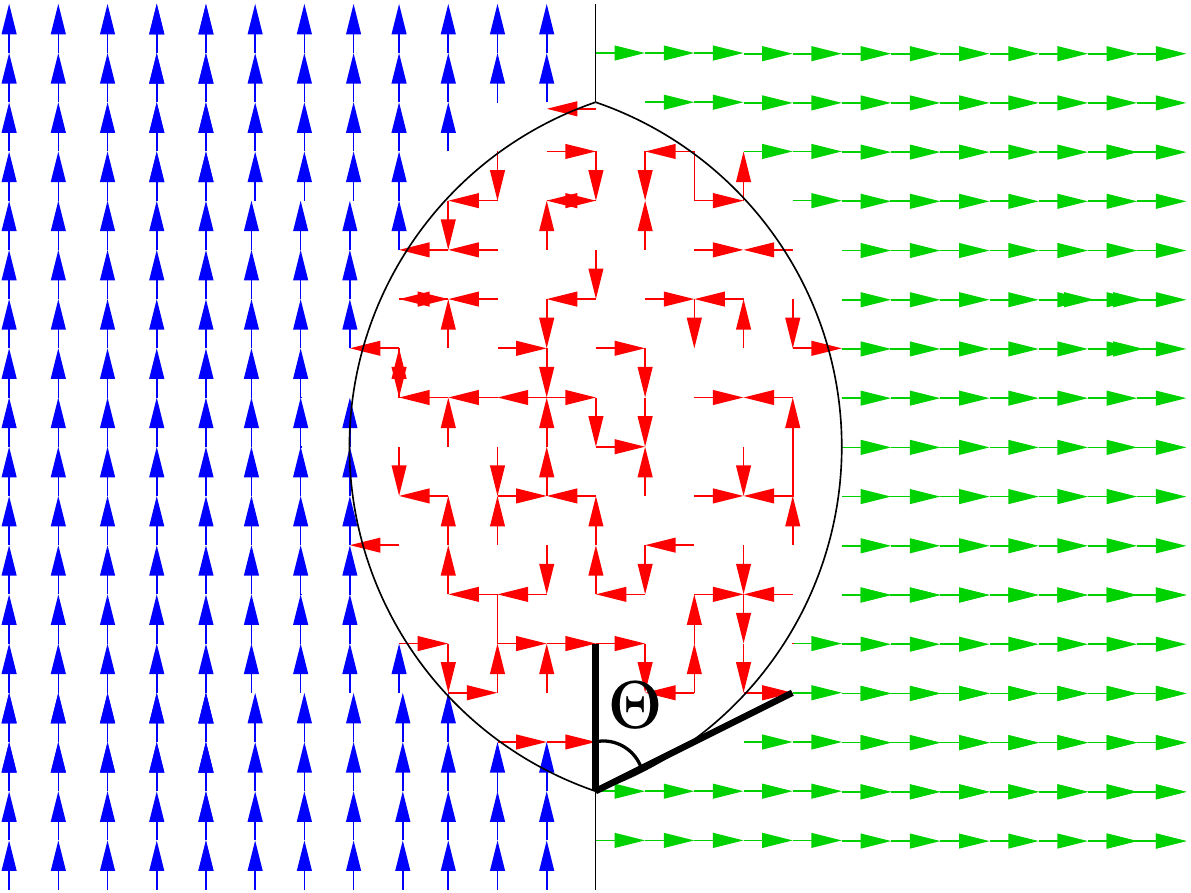}\hspace{20pt}
    \includegraphics[width=0.42\textwidth]{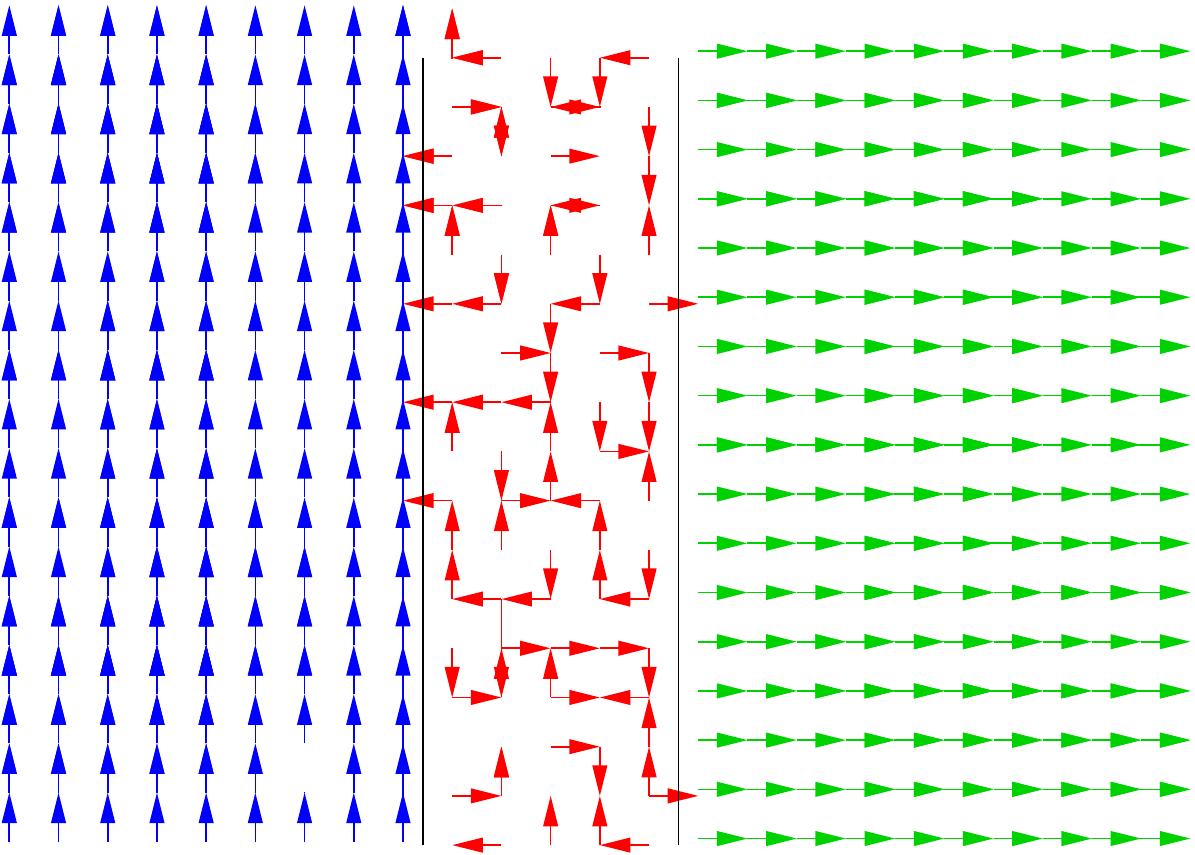}
    \caption{Illustration of wetting (left) and perfect wetting
      (right). In the case of perfect wetting the order-order interface is twice the order-disorder interface.\label{fig_wetting}}
  \end{center}
\end{figure}

The ratio of order-order to disorder-order interface tensions gives
the wetting angle
\begin{equation}
  \frac{\sigma_{\rm oo}}{2\sigma_{\rm od}} = \cos \theta, 
\end{equation}
where for perfect wetting $\theta=0$. See Fig.~\ref{fig_wetting}. Our result is a clear indication that perfect wetting holds for the three dimensional Potts model. This result has also been conjectured for the two-dimensional Potts model \cite{Borgs:1992qd}.

\acknowledgments
The work of B.L. is supported by the Royal Society through the University Research Fellowship. The authors acknowledge support from STFC under contract ST/G000506/1. The simulations discussed in this article have been performed on a cluster partially funded by STFC and by the Royal Society.


\begin{thebibliography}{99}
\bibitem{deForcrand:2000fi}
  P.~de Forcrand, M.~D'Elia, M.~Pepe,
  Phys.\ Rev.\ Lett.\  {\bf 86}, 1438 (2001).
  [hep-lat/0007034].

\bibitem{deForcrand:2005rg}
  P.~de Forcrand, B.~Lucini, D.~Noth,
  PoS {\bf LAT2005}, 323 (2006).
  [hep-lat/0510081].

\bibitem{Kajantie:1989iy}
  K.~Kajantie, L.~Karkkainen, K.~Rummukainen,
  Phys.\ Lett.\  {\bf B223}, 213 (1989).

\bibitem{Hietanen:2011jy}
  A.~Hietanen and B.~Lucini,
  to appear in Phys.\ Rev.\ E. [arXiv:1107.1637 [cond-mat.stat-mech]].

\bibitem{Wang:2001ab}
  F.~Wang and D.~P.~Landau,
  Phys.\ Rev.\ Lett. {\bf 86}, 2050 (2001).

\bibitem{Zhou:2005aa}
  C~Zhou and R~.N.~Bhatt,
  Phys.\ Rev.\ E {\bf 72}, 025701, (2005).

\bibitem{Morozov:2007aa}
  A.~Morozov and S.~Lin,
  Phys. Rev. E 76, 026701 (2007).

\bibitem{Morozov:2009aa}
  A.~Morozov and S.~Lin,
  J.\ Chem.\ Phys.\ {\bf 130}, 074903 (2009).

\bibitem{Bazavov:2008qg}
  A.~Bazavov, B.~A.~Berg, S.~Dubey,
  Nucl.\ Phys.\  {\bf B802}, 421-434 (2008).
  [arXiv:0804.1402 [hep-lat]].

\bibitem{MartinMayor:2006gx}
  V.~Martin-Mayor,
  Phys.\ Rev.\ Lett.\  {\bf 98}, 137207 (2007).
  [cond-mat/0611543 [cond-mat.stat-mech]].

\bibitem{Billo:2006zg}
  M.~Billo, M.~Caselle, L.~Ferro,
  JHEP {\bf 0602}, 070 (2006).
  [hep-th/0601191].

\bibitem{Caselle:2007yc}
  M.~Caselle, M.~Hasenbusch, M.~Panero,
  JHEP {\bf 0709}, 117 (2007).
  [arXiv:0707.0055 [hep-lat]].

\bibitem{Lee:1990ti}
  J.~Lee, J.~M.~Kosterlitz,
  Phys.\ Rev.\ Lett.\  {\bf 65}, 137-140 (1990).

\bibitem{Berg:1992qua}
  B.~A.~Berg, T.~Neuhaus,
  Phys.\ Rev.\ Lett.\  {\bf 68}, 9-12 (1992).
  [hep-lat/9202004].

\bibitem{Borgs:1992qd}
  C.~Borgs, W.~Janke,
  J.Phys.IFrance {\bf 2}, 2011 (1992).
\end{thebibliography}
\end{document}